# Semantic Data for Humanities and Social Sciences (SDHSS): an Ecosystem of CIDOC CRM Extensions for Research Data Production and Reuse

Francesco Beretta (Laboratoire de recherche historique Rhône-Alpes, CNRS / Université de Lyon)

## *Abstract*

Given the challenge of giant knowledge graphs created by major economic actors, which could virtually replace research in the Humanities and Social Sciences (HSS) in responding to public concerns, the question arises of how to increase the value of research data through their publication and networking, applying the FAIR principles. Both an epistemological and a semantic analysis show that the most relevant part of research data is factual information, understood as a representation of the objects observed by the scientific disciplines, their properties and their relationships.

This rich universe of information will be made understandable and therefore reusable through the application of foundational ontologies and a methodology based on the distinction between different levels of abstraction, allowing the collective development of one or more shared and reusable domain ontologies. This vision is being carried out around the *CIDOC CRM,* as core ontology, and *Semantic Data for Humanities and Social Sciences (SDHSS),* as a high-level extension of it, as well as an ecosystem of sub-domain extensions that can be easily managed through the *ontome.net* application. This will result in an interoperability that is semantically richer than the simple alignment of ontologies and less costly in terms of resources, and above all adapted to the scientific and humanistic project of the HSS.

## *Introduction*

The development over the last twenty years of methodologies and technologies of the semantic web and Linked Open Data (LOD) has made it possible to set up knowledge graphs of ever increasing size.[1] The creation of interconnected authority files[2], such as IdRef or VIAF[3] , or gazetteers such as Geonames or those produced by the Pelagios network[4], favours the integration of previously isolated data silos thanks to the identification and linking of information about resources, be these people, organisations, places, concepts, etc. The semantic web makes these resources and their properties accessible in the form of information whose meaning is formalised by ontologies so that it can be mobilised both by humans and by computers thanks to semantic reasoning or machine learning technologies[5]. Websites such as *data.bnf.fr* or *scienceplus.abes.fr* make bibliographic records and a rich universe of metadata accessible in data form.

The potential of this development has been recognised by search engines. They are improving the accuracy of their results through an artefact created in recent years, known as the giant knowledge graph. Thanks to advances in automated extraction of information from texts, it is now possible to envisage a rapid and almost unlimited supply of knowledge graphs. In 2020, Google's giant graph contained five billion entities and 500 billion "facts".[6] Researchers in the humanities and social sciences (HSS) cannot remain indifferent to this development, as these methods and technologies will not only affect the production of knowledge, but will also replace the HSS as providers of answers to questions of concern to civil society and the public.

By adopting these methodologies, the HSS can respond in at least two ways. Firstly, they will enable the full potential of the FAIR principles of "making data findable, accessible, interoperable and reusable"[7] to be realised. These principles, formulated by a group of scientists from the natural sciences and experts from the computer sciences, aim to promote the reuse of data generated by research in order to answer new questions.[8] Researchers are thus invited to publish not only the results of their investigations —the knowledge produced— but also to make available the data used to establish them.[9] Once the data published by HSS researchers are produced or at least made available in LOD formats and expressed according to a standardised ontology, it will be possible to construct one or more disciplinary knowledge graphs based on the information capital accumulated by research and thus the FAIR principles fully realised.

Secondly, given the importance of texts in several HSS disciplines the application of automated structured data extraction methods to written documents will make it possible to enrich information graphs with the content of texts and to make them "actionable" in a completely new way, revolutionising the way knowledge is produced. In other words, a paradigm shift is underway that is changing the methods of knowledge production and the learning of disciplinary tools.[10]

---

1 https://en.wikipedia.org/wiki/Knowledge_graph [all URLs were accessed on 3 February 2023]. This chapter is a revised and expanded version of my paper Francesco Beretta, Interopérabilité Des Données de La Recherche et Ontologies Fondationnelles : Un Écosystème d'extensions Du CIDOC CRM Pour Les Sciences Humaines et Sociales, in *Actes Des Journées Humanités Numériques et Web Sémantique*, ed. by Nicolas Lasolle, Olivier Bruneau, and Jean Lieber (Nancy, France, 2022), 2–22 <https://doi.org/10.5281/zenodo.7014341>.
2 https://help.oclc.org/Metadata_Services/Authority_records/Authorities_Format_and_indexes/Get_started/40Available_authority_files.
3 https://www.idref.fr/; https://viaf.org/.
4 https://www.geonames.org/; https://pelagios.org/.
5 Jens Dörpinghaus et al., Context Mining and Graph Queries on Giant Biomedical Knowledge Graphs, *Knowledge and Information Systems*, 64.5 (2022), 1239–62.
6 https://en.wikipedia.org/wiki/Google_Knowledge_Graph.
7 https://www.ccsd.cnrs.fr/principes-fair/. Cf. https://www.force11.org/group/fairgroup/fairprinciples.
8 "There is an urgent need to improve the infrastructure supporting the reuse of scholarly data", Mark D. Wilkinsonet al., The FAIR Guiding Principles for Scientific Data Management and Stewardship, *Scientific Data*, 3.1 (2016), 160018, Abstract, https://doi.org/10.1038/sdata.2016.18. Barend Mons et al., The FAIR Principles: First Generation Implementation Choices and Challenges, *Data Intelligence* 2.1-2 (2020), 1-9, https://doi.org/10.1162/dint_e_00023.
9 Cf. the journal *Scientific data* published by the Nature group : https://www.nature.com/sdata/, or the *Journal of Open Humanities Data* : https://openhumanitiesdata.metajnl.com/.

The condition for the realisation of this project is the adoption by the HSS disciplinary communities of ontologies and controlled vocabularies that are at the same time standardised, modular and extensible, allowing for a clearly defined common semantics that is flexible in its application. Indeed, it is important that the identity of the objects of scientific discourse, as well as the meaning of their properties and relationships, be clearly explained according to a sufficiently robust methodology, so that the data can both answer the precise questions of the researchers who produced them, and later be reused in the context of new research agendas. The challenge is thus both semantic and epistemological.

In order to reflect on the implications of this development for scientific methodology in HSS, we must first consider the content of the data to be shared and the relevance of the term 'knowledge graph'. In HSS, an important distinction must be made between information and knowledge: information can be defined as a representation of reality, and more precisely as a representation of the observed objects, their properties and relationships; knowledge as an interpretation of reality, an understanding of complex phenomena, their causes and their likely evolution.

It is true that semantic methods make it possible to derive new information from existing, which has led to such artefacts being called knowledge graphs. However, from the point of view of HSS, this is not knowledge in the true sense of the word, since knowledge requires, at the outset, the definition of a precise problem, a research question accompanied by lines of inquiry, and, at the end of the process, the creation in the minds of researchers of a model of reality, quantitative or qualitative, which is shared with a scientific community in order to be discussed and revised. This model will be proposed as the best available explanation, for the time being, of the structures, dynamics, causes and possible evolutions of the human and social world, past or present. From this perspective, the so-called knowledge graph is actually an information graph.

In this paper, I will first develop this last point by clarifying the epistemological distinction between information and knowledge, and between information and data, as it applies within the knowledge cycle in the historical sciences and, more generally, in HSS. A precise definition of these terms is essential in order to highlight the central issue of the application of ontologies in this field: it is indeed *information*, understood as a representation of the objects of scientific discourse, their properties and their relationships, that should be placed at the centre of data interoperability and the graph of the semantic web.

The second part will be dedicated to a presentation of a methodology allowing to collectively build a conceptualisation that is clearly defined, extensible and flexible enough to be applied to information modelling in different HSS domains. Given the diversity of information mobilised by the different disciplines, it is inconceivable to have a single ontology covering all domains: an intense dialogue is therefore needed between local conceptualisations, as produced by projects such as the present one, "Early Modern Professorial Career Patterns"[11], and a more abstract vision based on considerations and methodologies developed over the last decades in the field of research on foundational ontologies and semantic methodologies.[12] To support this approach, LARHRA has developed an online service, OntoME[13], which aims to manage and facilitate the modular and collaborative development of an ontology ecosystem adapted to the needs of HSS research.

---

10  Cf. Francesco Beretta, *Linked open data* et recherche historique : un changement de paradigme, Humanités numériques (accepted 16.12.2022, forthcoming).
11  https://pcp-on-web.de/ontology/0.2/index-en.html.

In the third part, I will propose a foundational analysis of the CIDOC CRM, a formal ontology standardised (ISO 21127:2014) and increasingly adopted in the field of HSS, designed for the integration of information from museums and cultural heritage conservation. The strengths and limitations of this ontology from an HSS perspective are highlighted, while a high-level extension is proposed, Semantic Data for Humanities and Social Sciences (SDHSS), whose aim is to promote the integration of conceptual models developed in research projects within an ontology ecosystem that enables straightforward interoperability of the data produced.

In the fourth part, I will illustrate some essential aspects of building a conceptualisation of social life, a central theme in HSS, and the condition for enabling information reuse in this research domain. In conclusion, I will point out the importance of the presented methodology to respond to the challenge for HSS research posed by semantic technologies and the growing importance of giant knowledge graphs.

## *The knowledge cycle in historical sciences*

Data, information, knowledge are polysemous terms that need to be carefully defined. Two diagrams summarising the process of knowledge production in the historical sciences from two different perspectives are useful here. The first is inspired by the stages of knowledge production formulated by Henri-Irénée Marrou in the form of a parabolic curve in a classic work on the "historian's craft"[14] (Fig. 1), as well as by the methodological steps of the social science research.[15] The choice of presenting this process in the form of a cycle underlines the iterative dimension of knowledge that is specific to the scientific approach and that also applies to the formulation and verification (or falsification) of hypotheses that is specific to the social sciences. The second diagram interprets, from the point of view of the historical sciences, the "data, information, knowledge" pyramid used by the information sciences to distinguish the different epistemic levels of knowledge[16] (Fig. 2). In this context, *knowledge production* is understood as a process, and *knowledge* as the content and result of the analysis and interpretation of information. This epistemological reflection models historical disciplinary practice, but is sufficiently general to be applicable, with the necessary adaptations, to other fields of HSS research.

---

12  https://en.wikipedia.org/wiki/Upper_ontology.
13  https://ontome.net/.
14  Henri Irénée Marrou, Comment Comprendre Le Métier d'Historien', in *L'histoire et Ses Méthodes*, ed. by Charles Samaran, Encyclopédie de La Pléiade (Paris: Editions Gallimard, 1961), 1465–1540: 1502.
15  Luc Van Campenhoudt et al., *Manuel de Recherche En Sciences Sociales*, 6th edn (Armand Colin, 2022).
16  Jennifer E. Rowley, The Wisdom Hierarchy: Representations of the DIKW Hierarchy, *Journal of Information Science*, 33.2 (2007), 163–80, https://doi.org/10.1177/0165551506070706.

As the diagram of the knowledge production cycle (Fig. 1) shows, all research must begin with the definition of a research agenda that fits within the horizon of existing knowledge, expressed in literature, and that defines the angle of approach of a subject of study, the methodology that will be adopted, and the general question. For example, in the context of the project on 'Career patterns of German professors from the sixteenth to the eighteenth century', the general question is about conditions that were necessary for professors to achieve professional success in the Early Modern university system. This general question will have to be translated into more specific ones, defining the lines of inquiry: for example modelling the careers of university teachers and their integration into scholarly networks, in articulation with the analysis of the content of their writings, possibly limiting the study to a region or a specific category. This first step is essential in order to be able to choose the sources to be used or the surveys to be carried out and to define the information to be gathered in order to answer the question.

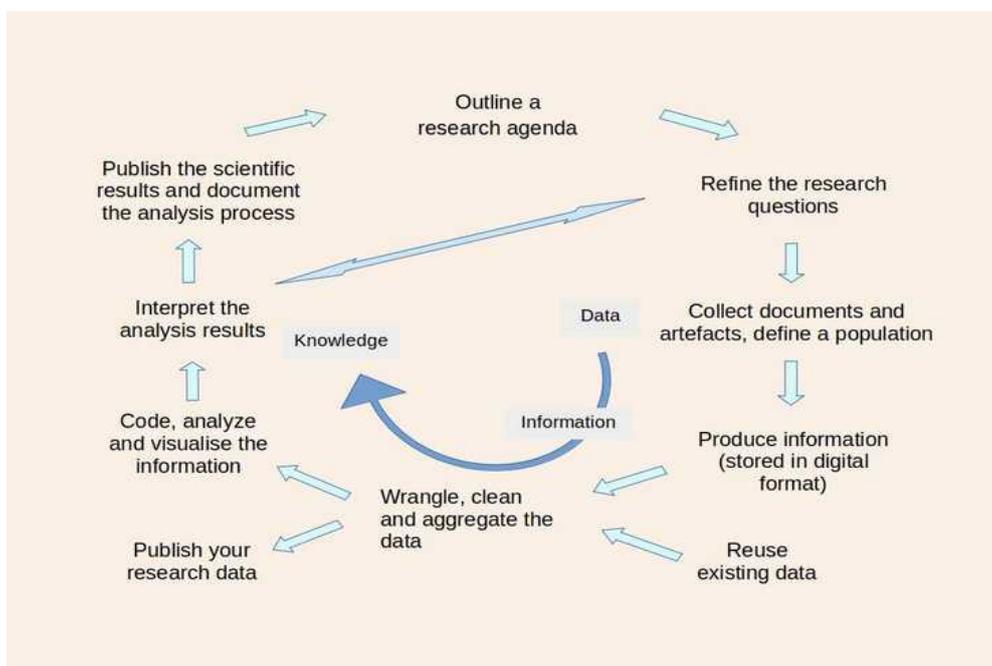

*Figure 1. Knowledge production cycle in historical sciences*

At this stage of research we are at the base of the pyramid (Fig. 2). It is important to note that data should be understood here in its primary and etymological sense, derived from the Latin *datum*, i.e. everything that is regarded as given and perceived as such by the observer, and not in the sense of digital data. By data we mean the observed reality as such, independent of the observer, be it directly observable in the social sciences or indirectly —through sources and physical remains— in the historical sciences. On the basis of their line of inquiry, HSS researchers must select from the mass of sources, or any other available and/or experimentally constructed trace of human activity, in order to gather the information that will be analysed and serve as a basis for knowledge. The questioning makes it possible to decide what information will be systematically retained and how it will be conceptualised and produced. This raises the issue of the conceptual model and the choice of digital storage technology, because while a spreadsheet may be adequate if one is limited to systematically collecting a certain number of characteristics of a population of individuals of the same type, as soon as one wishes to inform about complex relationships between different objects in space and time, it is essential to use a relational or graph-oriented database in order to capture the full wealth of information.

Let us note some of the initial results of this analysis. Information is at the heart of the scientific process. It can be defined as a representation of reality, and more precisely as an identification and representation of the objects in the world (people, organisations, artefacts, etc.), their characteristics (physical properties of objects, education and income levels of people, opinions, etc.) and their relationships in time and space (membership in organisations, exchange of messages or goods, journeys, etc.). Even if it is conceived from a representational perspective, and therefore with an explicit desire for objectivity in its production, information is always constructed, it is always the result of a question or a point of view. Consequently, research data, such as the contents of a spreadsheet or a database, are not 'data' in the primary sense, they do not immediately represent factuality, because they always presuppose a specific questioning and conceptualisation that allowed their collection. It is therefore essential to make explicit the semantic content of digital data and the way in which it has been produced, as an indispensable condition for its reuse.

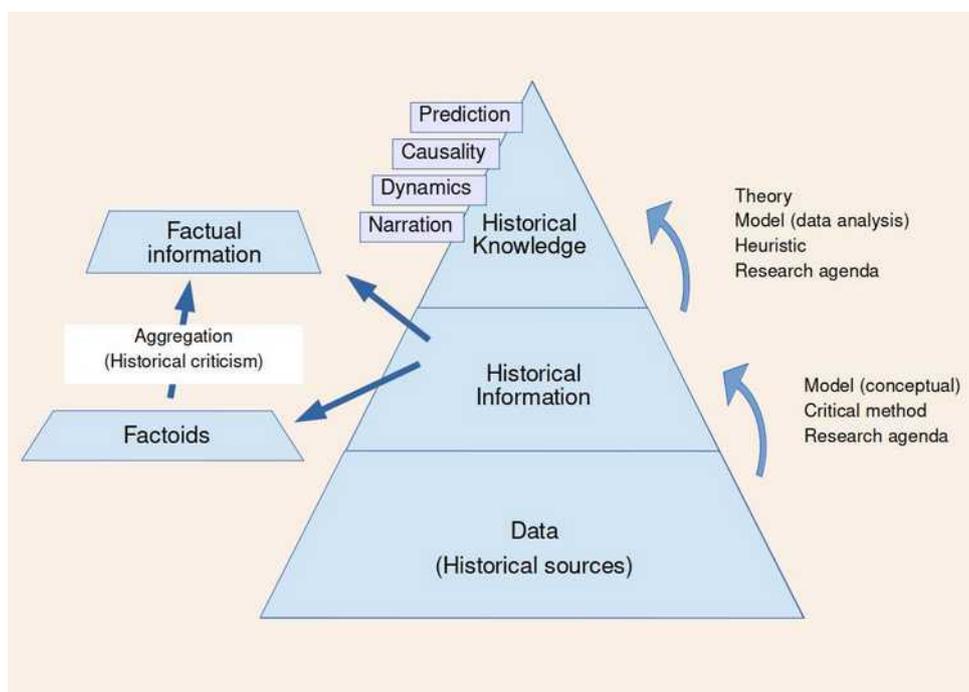

*Figure 2. Pyramid « sources, information, knowledge » as interpreted in historical sciences*

Let us also note that, in the pyramid, information is articulated on two levels: one can aim at a faithful reproduction of the content of the sources, or at a daily observation of economic transactions or of the manifestations of contemporary social relations, situating oneself on an epistemic level that is generally called that of *factoids*.[17] In this scenario, we will have access to extensive but redundant, or even contradictory information about the same properties of objects.

---

17 Michele Pasin, John Bradley, Factoid-Based Prosopography and Computer Ontologies: Towards an Integrated Approach', *Literary and Linguistic Computing*, 30.1 (2015), 86–97, https://doi.org/10.1093/llc/fqt037.

Taken as such, this information will inevitably distort the results of the analyses. As a matter of fact, if we want to compare, to return to our example, the careers of university teachers, it is not enough to collect the multiple mentions in different sources of the same career stages of the same people, but it is necessary to aggregate them in order to identify and reconstruct the career segments of each person. In the event of disagreement between sources, it will therefore be necessary to make choices so that the analysis is not distorted by the redundancy of the facts and the information produced is the best possible approximation to the factuality of the characteristics and relationships of the objects studied. Good quality factual information is generally the indispensable basis for knowledge production. The epistemic levels of factoids and factual information are therefore fundamentally different.

Once this aggregation has been carried out, the information must be coded and simplified according to the lines of inquiry. It is at the level of this second aggregation and modelling operation that the questioning is injected, in order to be able to apply to the information, suitably prepared, standardised and coded, a panoply of tools: statistical software, network analysis, representation and spatial analysis, etc. (Fig. 1). The model, in the statistical sense, that emerges from these analyses has an eminently heuristic function, because the mathematical and visual representations produced by analysis software always require critical discussion, contextualisation and interpretation. At the same time, the analysis software makes it possible to make visible significant phenomena that would otherwise be impossible to see "with the naked eye" —for example, the comparison of career segments and the identification of recurrent prosopographic profiles among hundreds of teachers over several centuries, in relation to their distribution in geographical space— despite the considerable volume and complexity of the available information, which is conveniently condensed and simplified thanks to the groupings and codings used in this phase of the analysis.

At the end of this process, researchers produce knowledge as an answer to the general questions of their research agenda and publish the results of their investigations. It clearly appears from these two diagrams that there is an essential epistemic distinction between the knowledge produced in this way and the information on which it is based, since the (hypo)theses to which knowledge leads — creating a model of the world intended as a representation of the complex dynamics of phenomena, their structures or their causes— always involve a synthesis of information and an interpretation that goes beyond the simple representation of factual reality. Therefore, in the logic of open science, it is essential to publish not only the knowledge obtained, but also the research data themselves, i.e. the information collected and analysed, in order to facilitate the verification of the hypotheses put forward by exposing them to "falsification" in the logic of a reproducible scientific approach.[18]

This analysis shows the full potential of the new digital methodologies in view of knowledge production in HSS, as it is now possible to go well beyond the volume of data that individual researchers can collect, and to access increasingly rich and voluminous pools of information. At the same time, two principles emerge that need to be rigorously applied to enable the reuse of research data. On the one hand, the information collected and expressed in digital data should, as far as possible, be conceived as a representation of factual reality as such, avoiding as far as possible biases introduced by research perspective or data coding. The aggregation and simplification that precede the analysis should therefore take place in a second phase, while the sharing of the data will mainly concern the information collected in the first phase of the research. On the other hand, the data to be shared must be produced with clearly defined semantics in order to make the meaning of the information explicit and to allow its reuse. Moreover, this process must be carefully documented in order to allow other researchers to identify possible shortcuts introduced in the conceptual model and to have indications of the data quality, providing sufficiently rich metadata about information production[19].

---

18  https://en.wikipedia.org/wiki/Falsifiability.

*Foundational Ontologies and Ontology Engineering Methodology*

A key question remains behind the scepticism often expressed about the real possibility of reusing data produced by HSS for new research: if information is - as we have shown - the product of a conceptual construction, resulting from the application of a research question and adopting a conceptualisation in relation to the domain, is this not a major and quasi-structural obstacle to the reuse of data? Is a representation of factual reality through information really possible, or at least expressible in the form of interoperable data? This apparent difficulty explains the decision of the "Early Modern Professorial Career Patterns" project to limit itself to the development of a domain ontology dedicated to the study of the research agenda specific to the project, without attempting a more generic modelling and alignemnent of core classes with existing standards.[20] This is a common practice in HSS and an obstacle to the reuse of data advocated by the FAIR principles.

There is, however, a positive answer to this fundamental question, and it is provided by several decades of publications in the field of foundational ontologies and knowledge engineering methodologies. One of the main players in this discipline, Giancarlo Guizzardi, writes in a critical and stimulating article that information interoperability is only possible if we adopt "formal, shared and explicit representations of conceptualisations, or what the field of knowledge representation has conventionally called ontologies". And this author specifies that it is not the fact of expressing the conceptual model of a particular project thanks to formal logic or the *Ontology Web Language (OWL)* that creates an ontology, but rather the fact of carrying out an analysis of the essential aspects of reality, such as the identity of the objects, their relationships, their compositions and dependencies, and this by adopting a high-level conceptualisation that is transdisciplinary and can be applied to several fields of scientific discourse. This is the role of foundational ontologies[21], a field in which Guizzardi is active as one of the creators of the *Unified Foundational Ontology* (UFO).[22]

---

19 For the field of historical sciences see for example the *Historical Context Ontology* (HiCO), based on the PROV-O digital data origin documentation ontology, https://marilenadaquino.github.io/hico/. In a more technological dimension, see the new RDF-Star specification which is increasingly implemented in semantic databases, https://w3c.github.io/rdf-star/cg-spec/2021-12-17.html.
20 Cf. Introduction, https://pcp-on-web.de/ontology/0.2/index-en.html
21 Giancarlo Guizzardi, Ontology, Ontologies and the "I" of FAIR, *Data Intelligence*, 2.1–2 (2020), 181–91, https://doi.org/10.1162/dint_a_00040.
22 Giancarlo Guizzardi et al., UFO: Unified Foundational Ontology, *Applied Ontology*, 17.1 (2022) https://doi.org/10.3233/AO-210256.

Looking at the classes and properties produced in the ontology of the "Early Modern Professional Career Patterns" project, several issues can be observed. The choice of the class *pcp:Lecturer* —"university lecturer at early modern universities [...] whose career patterns are the focus of the research project relevant to the given ontology"— as the pivot of the ontology raises the problem of its articulation with the class *pcp:Person*, because while the latter is rigid in the sense of the OntoClean methodology (see below), the former is not, and represents only a temporary condition or social role of a person, just like the class *pcp:Student*, and not an essential property of the instances of that class.

But is it then legitimate to associate another basic class in the ontology, *pcp:StageOfLife*, understood as phenomena situated in time and physical space, with *pcp:Lecturer* and not with *pcp:Person*? And thus people who do not teach do not have life segments or births or deaths? In fact, even the class *pcp:Person* has a whole series of properties, which are inherited by the subclass *pcp:Lecturer*, but they are expressed in the form of simple relations, not classes; birth and death date, kinship relations, etc. These modelling choices, which introduce redundancy and inconsistency in the conceptualisation of the same information, can be explained by the focus on a specific population, in the context of a specific research project, but the consequence is that the model is biased, making interoperability and reuse of data more difficult, and unduly limiting the expressiveness of the ontology.

There are also more subtle issues, such as the distinction between curricula, represented by *pcp:AcademicDocuments*, which are propositional objects formulated by the sources —for example, the yearly planned course outline— and the actual teaching that took place, which appear as a subclass of *pcp:AcademicOffice* and thus as a descendant class of *pcp:StageOfLife*, representing teaching activities as spatio-temporal events. But if the aim is to inform career segments, could we not simply list information about people's roles or status (teacher, lecturer, academy member) as social qualities, without expressing the teaching activity as such, which may sometimes not take place due to illness or other reasons? And if so, what kind of conceptualisation should be used for this purpose?

A recent issue of the journal *Applied Ontology* provides an instructive illustration of the approach to be taken.[23] The authors of the main foundational ontologies, *Basic Formal Ontology* (BFO), *Descriptive Ontology for Linguistic and Cognitive Engineering* (DOLCE), *A Top Level Ontology within Standards* (TUpper), which make up the ISO 21838 standard, as well as UFO and a few others, were invited to propose, from the point of view of their ontological analysis, the modelling of some classical knowledge engineering questions concerning the description of artefacts and their components, the changes in the properties of objects over time, or the representation of modification of social situations. The aim is to enable semantic engineers to understand the philosophical underpinnings of the main foundational ontologies —as they have different emphases and are based on different philosophical approaches— and the specificities of their conceptualisations, in order to choose the one that seems to be the most efficient in terms of foundational analysis of the domain concerned.

---

23  Stefano Borgo et al., Foundational Ontologies in Action, *Applied Ontology* 17.1 (2022), https://doi.org/10.3233/AO-220265.

Among these ontologies, DOLCE is particularly well adapted to the HSS perspective and is frequently used in this field.[24] We have chosen to use it as a reference for our foundational analysis, even though other ontologies —in particular UFO with the UFO-C2 module— also offer interesting analytical perspectives for modelling social phenomena. DOLCE is an ontology of particulars, i.e. it does not aim to identify the metaphysical substance of reality, but "to make explicit already existing conceptualisations through the use of categories whose structure is influenced by natural language, the structure of human cognition and social practices". This ontology is therefore particularly well suited to the programme of creating an interoperable conceptualisation of information in HSS presented above.

Moreover, DOLCE has been complemented not only by some extensions modelling roles and artefacts, and even social and cognitive aspects, but above all by the sister ontology *Descriptions & Situations* (D&S), developed in the same original project, whose domain is the foundational modelling of different perspectives of agents on the same world events.[25] The notion of situation is defined as an interpretation of events based on a particular conceptualisation, i.e. representations shared by agents and expressed by a description that assigns specific roles and connotations to the participants in the event. D&S was integrated with DOLCE to produce the *DOLCE Lite Plus* (DLP) ontology, which we use as a reference for our analytical work, and which was also reformulated and simplified in *DOLCE Ultra Light* (DUL), the base of the modelling approach of *Ontology Design Patterns*.[26]

If DLP provides an ontological basis for distinguishing between events in the world and the representations developed by different actors about these same events, this approach has also made it possible to model the activity of scientific communities from a constructivist point of view[27], in a way that is consistent with the distinction of different epistemic layers presented above using the DIK pyramid model, and responds to the challenge of reconciling a transdisciplinary conceptualisation of information (DOLCE) with the specificities of each discipline (D&S). In other words, the conceptuality of DLP can be applied to HSS in order to epistemologically ground the method of data interoperability proposed here: the same factuality expressed by the information can correspond to different epistemological "situations", i.e. different interpretations according to the points of view of different disciplines, producing different knowledge. However, these interpretations should only be used as an overlay to the production of information, i.e. only in a second phase of the research, when the data are aggregated and coded in order to analyse them and answer the research question, as we have seen. At the same time, it is important to model the information gathered during the research as objectively as possible, i.e. as independent of a research agenda as possible.

---

24  Descriptive Ontology for Linguistic and Cognitive Engineering (DOLCE), *WonderWeb Deliverable D18* (Laboratory for Applied Ontology: Trento, 2003). Stefano Borgo et Masolo Claudio, Foundational Choices in DOLCE, in *S. Staab and R. Studer (Eds.), Handbook on Ontologies,* (Berlin / Heidelberg: Springer-Verlag, 2009), 361–81.
25  C. Masolo, S. Borgo, A. Gangemi, N. Guarino and A. Oltramari, WonderWeb deliverable D18 ontology library (final), Laboratory for Applied Ontology, Trento, 2003.
26  Aldo Gangemi, Valentina Presutti, Dolce+D&S Ultralite and Its Main Ontology Design Patterns, in *Ontology Engineering with Ontology Design Patterns: Foundations and Applications,* Pascal Hitzler et al. *(eds)* (Amsterdam Berlin: IOS Press AKA, 2016), 81-103.
27  Gangemi Aldo, Norms and plans as unification criteria for social collectives, *Autonomous Agents and Multi-Agent Systems* 17.1 2008, 70-112, https://doi.org/10.1007/s10458-008-9038-9.

DOLCE thus proposes a conceptualisation —valid at least in the context of our civilisation— that allows transdisciplinarity in the production of information. It should be noted that this conceptualisation has been carried out using the OntoClean methodology developed by Nicola Guarino and Emil Welty, whose aim is to formalise foundational analysis using philosophically fundamental categories such as essence (and rigidity), identity, unity and dependence.[28] Therefore, by using DOLCE to analyse the conceptualisation of a domain, one is already on the right track to define a robust and interoperable ontology, avoiding a number of modelling biases.

DOLCE divides particulars, i.e. the entities to which scientific discourse refers, into four distinct and non-intersecting classes: endurants, perdurants, qualities and abstracts. The essential difference between endurants and perdurants is their relationship with time: endurants preserve their identity through time, even if their properties evolve; perdurants, which develop in time, and with time, are at each moment only partially present, although identifiable as a whole. Endurants and perdurants are linked by the relation of *participation* of the former in the latter, for example the participation of people in a meeting or a battle. We may add, and it should be noted, that perdurants, as spatio-temporal phenomena, would have virtually no observable existence if they had no participants. Endurants are therefore *identifying components* of perdurants: for example, the birth of a person is not a birth in general, it is the birth of that person, from which it is inseparable.

There is then a distinction between dependent and independent objects, because a hole in a shirt does not exist without the shirt, nor does a cave exist without the mountain (these are features), and the material that makes up a table (the wood, amount of matter) has an identity that is different from the one of the table itself, the latter resulting from its form (physical object). In the sphere of conceptual objects we have mental and social objects, and in particular roles and collectives, which result from the notion of classification and are analysed in the extensions of DOLCE.[29]

Two other classes, qualities and abstracts, provide a complete articulation of the whole of human discourse. *Qualities* are observable properties of endurants or perdurants that are specific to them. These include occupied space as a property of physical objects, while temporality is a property specific to events. Note that in DOLCE qualities are conceived as inherent to objects: each chair has its own colour at a given moment. Each instance of the quality colour will therefore have its own value, i.e. it will occupy a point or "region" in a reference space, which is expressed by the notion of region as a subclass of the ontology's class abstracts. *Abstracts* are discourse entities which, having no temporal or spatial properties of their own, nor the status of qualities, are situated outside observable entities and, it may be added, appear to be the product of research community conventions —metric measures, for example— which allow property values to be located in a reference space. Other foundational ontologies locate these 'abstracts' as subcategories of artefacts. From an epistemological point of view, it is important to note the distinction clearly visible in DOLCE between phenomena on the one hand and abstract reference spaces on the other, for example geographical places versus coordinates in the WGS84 reference frame, which allow phyisical places to be situated in the abstract space of the Earth's reference geoid.

---

28   Nicola Guarino, Christopher A. Welty, An Overview of OntoClean, in: *Handbook on Ontologies* (see fn. 24).
29   E.g. Claudio Masolo et al., Social Roles and Their Descriptions, in *Principles of Knowledge Representation and Reasoning: Proceedings of the Ninth International Conference (KR2004), Whistler (Canada),* 2004, 267–77, http://www.aaai.org/Library/KR/2004/kr04-029.php ; Daniele Porello et al., The Ontology of Group Agency, in *Formal Ontology in Information Systems - Proceedings of the Eighth International Conference, FOIS 2014,* Rio de Janeiro (Brazil), 2014, 183–96, https://doi.org/10.3233/978-1-61499-438-1-183.

If we apply these categories to the analysis of information as a representation of factual reality, we find in these four classes the essential elements introduced earlier: the objects represented by the information are the endurants (persons, artefacts, groups, etc.), their properties are expressed by qualities (colour, weight, size, etc.) located in the reference spaces specific to the different disciplines, while their relations and their evolution in time are captured thanks to their participation in the perdurants. As far as their situation in physical space is concerned, in DOLCE this is conceptualised as the quality of the endurants and is therefore only indirect for the perdurants, whose projection in physical space corresponds to that of the agents involved in the events. We thus have the conceptual tools necessary to build interoperable domain ontologies. Indeed, it should be noted that the categories presented are independent of specific scientific theories or problems. Therefore, when factual information is captured by adopting this conceptualisation, it will allow the properties and relationships of objects to be reproduced in the form of data in the most objective way possible, while leaving to the scientific disciplines the task of explaining and interpreting these same properties and relationships.

Since foundational ontologies propose conceptual handles but are not intended to be used directly, it is necessary at this stage to develop a domain ontology, i.e. a conceptualisation of a particular domain of scientific discourse.[30] This process could be carried out directly from one's own specific research model, evaluated against the foundational categories, which would already allow for interoperability. However, in order to facilitate ontology design and to promote interoperability of data produced by scientific disciplines, it seems more useful to proceed with a methodology of *abstraction layers* (Fig. 3, left side).

---

30  Borgo Stefano, Foundational ontologies in action (see fn. 23).

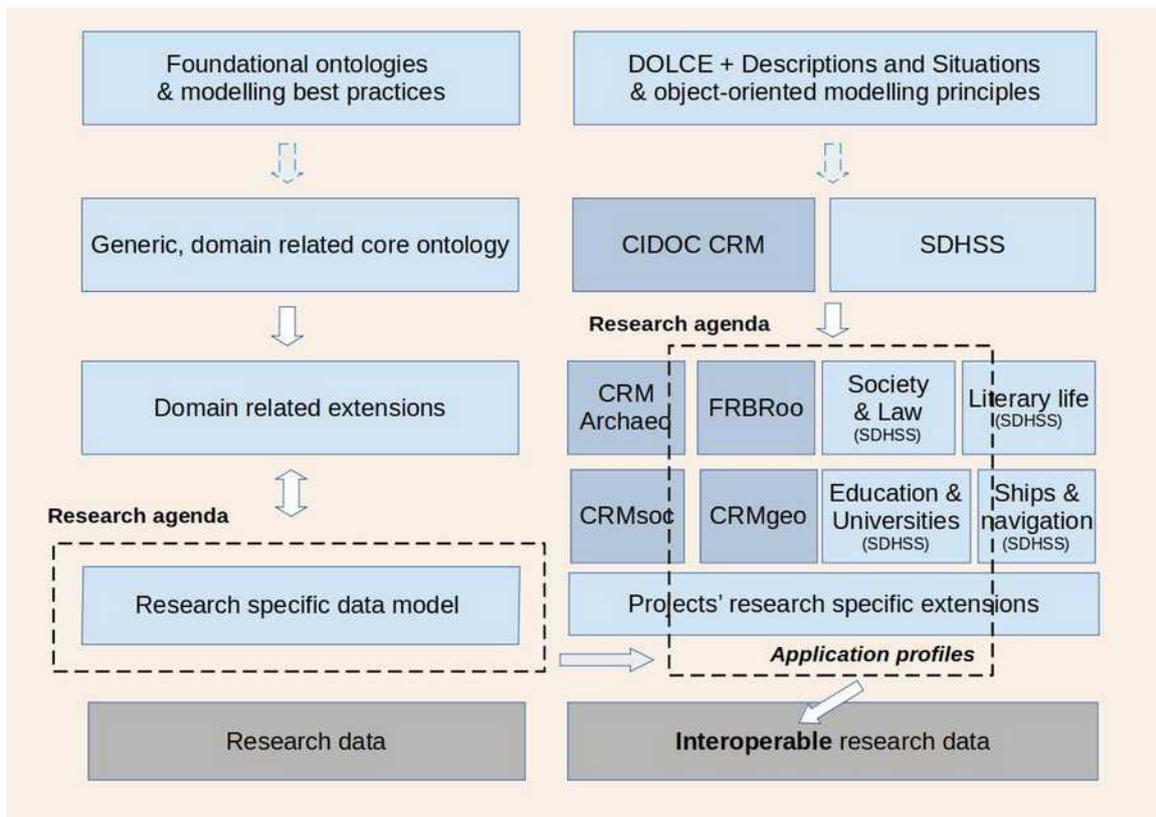

Figure 3. *Methodology for ontology development using multiple levels of abstraction*

This method involves adopting a high-level domain ontology, a core ontology, which provides the basic classes and properties for describing the objects studied by the discipline in question. This conceptualisation needs to be verified against the classes of a foundational ontology in order to improve its quality and expressiveness. These classes, such as perdurants or endurants, are therefore not used directly, but serve as a reference. Then, one can develop extensions of the core ontology by subdomains within the discipline, e.g. economic or social history for historical sciences, proposing classes and properties that capture specific information. Finally, one chooses among the already existing classes and properties those to be used in one's own project, using application profiles intended as suitable collections of classes and properties. If necessary in order to process the information at hand one can add those that are still missing. The advantage of this layered approach is, on the one hand, that it avoids having to reinvent a domain conceptualisation for each new project. On the other hand, the reuse of existing, clearly defined classes and properties greatly facilitates interoperability. Logically, this requires strict respect for the existing conceptualisation, i.e. an understanding of the "intension" of classes and properties, thereby ensuring interoperability thanks to a formalised and shared semantics.

## A core extension of CIDOC CRM : Semantic Data for Humanities and Social Sciences (SDHSS)

The usefulness and importance of this layered methodology became clear in the evolution of the *symogih.org* project towards semantic web methodologies and technologies. I will briefly illustrate the stages of this process, as they explain the choice of proposing CIDOC CRM as the core ontology for HSS, as well as the need to add a high-level extension integrating some aspects more specifically related to the information processed by these disciplines, especially in relation to the question of conceptualising social life.

The *symogih.org* project, "Modular system for the management of historical information", was born in 2008 from the desire of a number of historians at the Rhône-Alpes Historical Research Laboratory (LARHRA) in Lyon to pool the structured data generated by their research in order to enable its reuse.[31] About fifty individual or collective projects have used the collaborative Virtual Research Environment (VRE) created by this project. The interoperability of information in the VRE has been achieved through the creation of a generic and open conceptual model, abstract enough to meet different information production needs, extensible, shared within the VRE and published on the main project website.[32] The meaning of the data, i.e. the semantics of the information they carry, is thus made explicit, allowing it to be easily and consistently reused.

In 2016, during a process of ontology formalisation aimed at bringing the *symogih.org* project into the realm of semantic interoperability, LODs and FAIR principles[33], it seemed useful to integrate into the context of the CIDOC CRM the experience gained so far. This conceptual model, which obtained the status of an ISO standard in 2006, models the museum domain and therefore has important intersections with the historical research domain. Moreover, the CRM development methodology, which is object-oriented and follows conceptual principles partly similar to those applied by OntoClean, provides a coherent system of high-level classes articulated in a hierarchy of property inheritance based on a fine-grained analysis of the relationships between objects and events.[34] Because of this genericity, it seemed appropriate to adopt it as the core ontology for the historical sciences and, more broadly, for HSS.

Over time, however, the difficulty of aligning all the information already modelled in the *symogih.org* project with the CIDOC CRM classes became apparent, even taking into account its family of extensions. In the light of the semantic methodologies presented above, the reasons for this difficulty are clear. On the one hand, while there is certainly an intersection of domains, there remains a significant difference between the purpose of CIDOC CRM, i.e. the integration of museum data through a process of ontological abstraction, and the domain of HSS discourse, which is much broader and requires nuance, complementation and specialisation. The implementation of an abstraction layer methodology therefore appears appropriate. On the other hand, a fundamental analysis of the CIDOC CRM, as well as the application of the OntoClean method, allows to highlight some incompatible aspects of the respective conceptualisations, beyond an apparent homonymy of classes.

---

31  http://symogih.org. Francesco Beretta, Pierre Vernus, Le projet SyMoGIH et la modélisation de l'information : une opération scientifique au service de l'histoire, *Les Carnets du LARHRA* 1 2012, 81-107, https://shs.hal.science/halshs-00677658.
32  http://symogih.org/?q=type-of-knowledge-unit-classes-tree.
33  Francesco Beretta, L'interopérabilité des données historiques et la question du modèle : l'ontologie du projet SyMoGIH, in *Enjeux numériques pour les médiations scientifiques et culturelles du passé,* Brigitte Juanals and Jean-Luc Minel (eds) (Paris: Presses universitaires de Paris Nanterre 2017), 87–217, https://halshs.archives-ouvertes.fr/halshs-01559816.
34  Cf. http://www.cidoc-crm.org/.

The adoption of DOLCE Lite Plus as a foundation layer (Fig. 3, right) has made it possible to clarify the issues at stake and to identify the aspects that are not modelled in the CRM, or at least not in a fully satisfactory way from the point of view of HSS.[35] It is therefore essential to add, at the same level as the core CRM ontology, an extension that we have called Semantic Data for Humanities and Social Sciences (SDHSS), which enriches the high-level ontology with some classes that are essential for structuring the whole domain. And also to fill the gaps in the sub-domains, such as social and economic life, with lower-level extensions, which can only be done by creating an ecosystem of extensions that will be progressively enriched as projects require. We hope that the development of this ecosystem will become increasingly participatory, allowing a large number of HSS projects to challenge the conceptualisations proposed, testing them in their research, and to progressively build a true semantic interoperability of data.

This data integration project, based on robust and shared semantics, motivated the creation of the *Data for History* consortium[36], which was founded in November 2017 during a workshop organised at the École Normale Supérieure in Lyon, followed by a second workshop in Lyon in 2018, then a meeting in Leipzig in 2019 and the first (online) international conference in May-June 2021, organised by the Chair of Digital History at the Humboldt University in Berlin[37], and which is currently being continued by the *Data for History Lectures.*[38]

This objective also justified the creation of an indispensable online support: the collaborative ontology design and management application implemented by LARHRA since 2017, OntoME (Ontology Management Environment)[39], which has been adopted by several projects.[40] OntoME allows the handling of multiple namespaces with autonomous rights management per project, the import and export of ontologies in RDFS and OWL-DL, the creation of application profiles to be used in data production VREs such as *geovistory.org*. OntoME also allows the creation of subdomain specific extensions, such as those of the *Maritime History*[41] or the French ANR funded *Processetti* project[42], adapted to the information production needs of the respective research agenda, but developed on the basis of the abstraction layer methodology presented above. The lifecycle of these extensions can be limited to the duration of the project, or they can be reused and completed

---

35 Beretta Francesco, « A challenge for historical research: Making data FAIR using a collaborative ontology management environment (OntoME) », *Semantic Web* 12(2021)2, pp. 279-294, https://doi.org/10.3233/SW-200416.
36 http://dataforhistory.org/. Francesco Beretta, Vincent Alamercery, Du projet symogih.org au consortium Data for History - La modélisation collaborative de l'information au service de la production de données géo-historiques et de l'interopérabilité dans le web sémantique, *Revue ouverte d'ingénierie des systèmes d'information*, 1.3 (2020) https://doi.org/10.21494/ISTE.OP.2020.0532.
37 http://dataforhistory.org/3rd-data-for-history; https://d4h2020.sciencesconf.org/.
38 http://dataforhistory.org/news
39 https://ontome.net.
40 In particular, two EU-funded projects used OntoME for the preparation of the data model: Silknow (ERC project) et Read-it (JPICH project).
41 https://ontome.net/namespace/66 . Francesco Beretta et al., Geohistorical FAIR data: data integration and Interoperability using the OntoME platform, in: *Time Machine Conference 2019*, Dresden (Germany), https://shs.hal.science/halshs-02314003v1.
42 https://ontome.net/profile/15.

by new projects working on the same subdomains, in the logic of a dynamic and evolving ecosystem.

The necessary extension of the CIDOC CRM (hereafter CRM) with a core ontology of the same abstraction level, adapted to the information production needs of the HSS, will seek to respect as much as possible the conceptualisation of the CRM in order to ensure the greatest compatibility according to the principles expressed in the standard.[43] The aim of this approach is to express the information produced as a representation of objects, their properties and their relationships, with the greatest possible objectivity and rigour. The question of how to conceptualise the production of information and how to express its quality in relation to the sources from which it is extracted, which is also essential for interoperability, will not be addressed here, especially as it has already given rise to a number of proposed solutions, such as the Historical Context Ontology (HiCO)[44], an extension of PROV-O.[45]

This undertaking must start with an analysis of the CRM in the light of the OntoClean methodology and the foundational ontologies, in our case DOLCE. This study has already been carried out and has highlighted a range of issues in the CRM, with proposals for improving the formalisation of the ontology, some aspects of which I will mention in the following pages.[46] The structure of the ontology can be discovered by inspecting the tree of classes published in OntoME.[47] By progressively unfolding the tree and browsing its branches, one will find the classes I am going to present and will be able to access the definition of their "intension" in the scope notes as well as those of their properties. In the tree, without the need to login, we will find, in addition to the CRM and FRBRoo, the namespaces that are part of the SDHSS project. To distinguish them, I will prefix the classes and properties with *crm* for the CRM and *sdh* for the new high-level extension.

The root class, *crm:E1 Entity*, contains all the objects in the CRM discourse domain. Note that the values, literal values in the sense of RDF, are not part of it and are collected in the class *crm:E59 Primitive Value*. They are therefore outside the ontology, which refers to existing standards for expressing these values. If we unfold the tree, we notice the two essential classes *crm:E77 Persistent Item* and *crm:E2 Temporal Entity*, corresponding respectively, at least at first glance, to the classes Endurant and Perdurant of DOLCE. Missing are the classes Quality and Abstract, while there are four other root level classes (*crm:E54 Dimension*, *crm:E53 Place*, *crm:E52 Time Span*, *crm:E92 Spacetime Volume*). These are, in the light of the DOLCE conceptualisation, regions and therefore subclasses of Abstract, as they correspond to a particular position in a conventional reference space. They are therefore grouped in the extension's *sdh:C5 Abstract Region* class to emphasise this analysis and to avoid confusion.

In this respect, it should be noted that there is a widespread misunderstanding in projects that use the class *crm:E53 Place* to model geographical places: according to the CRM, a place is a pure extent in a reference space and should therefore be more properly called S*pace* and not *Place*, which is confusing. This is confirmed by the fact that according to the CRM one can take a picture of an instance of *crm:E27 Site* —a subclass of *crm:E26 Physical Feature* commonly used for modelling archaeological sites— but not an instance of *crm:E53 Place* whose ontological substance is supposed to be a part of an abstract reference space.[48] The class *sdh:C13 Geographical Place* has therefore been added as a subclass of *crm:E26 Physical Feature* in the SDHSS extension to clarify the distinction in the ontological substance between a physical and an abstract space, and to take into account the fact that a geographical place can be projected over time into different instances of *crm:E53 Place*, such as a city or a territory, whose projection in space evolves over the years.

---

43  https://cidoc-crm.org/Version/version-7.1.2, "Extensions of CIDOC CRM", xiv-xvi.
44  https://marilenadaquino.github.io/hico/
45  https://www.w3.org/TR/prov-o/
46  Sanfilippo Emilio M., Beatrice Markhoff, Ontological Analysis and Modularization of CIDOC-CRM, *Formal Ontology in Information Systems*, 2020, 107-121, https://doi.org/10.3233/FAIA200664.
47  https://ontome.net/classes-tree.

As far as the class *crm:E77 Persistent Item* and its subclasses are concerned, they express a conceptualisation not very far from that of DOLCE and include independent objects and their associated features, physical objects and their non-material counterparts. However, there are some peculiarities that have been pointed out as not conforming to the OntoClean method. First, a distinction between agent (*crm:E29 Actor*) and "inert" object (*crm:E70 Thing*) that is based more on intentionality than on a more objective classification, with actors being persons, "individually or in groups, who have the potential to perform intentional actions". Animals and non-human agents are thus excluded from the *crm:E29 Actor* class and are modelled in the form of *crm:E24 Physical Man-Made Thing* or *crm:E20 Biological Object* instances, further down in the hierarchy, but we are surprised to find again, at this level of the taxonomy, humans, here understood in their biological materiality, or "animality". DOLCE's taxonomy is much stricter in terms of the OntoClean method.

A certain sense of ontological 'ambiguity' is also apparent in the definition of *crm:E72 Legal Object* class, which is distinct in the tree hierarchy from the *crm:E71 Man-Made Thing class*, although the *crm:E24 Physical Man-Made Thing* class appears then lower in the hierarchy as a subclass of the two preceding classes. The function of the class *crm:E72 Legal Object* is to group together objects over which a right belonging to the actors can be exercised, expressed by the *crm:E30 Right* class. It has been rightly pointed out that this class is therefore anti-rigid in the OntoClean sense, i.e. being subject to ownership or other rights is certainly possible, but not essential for the definition of the class, which would invite the removal of *crm:E72 Legal Object* from the class hierarchy of persistent items and the expression of this legal connotation with a time-indexed classification relation as used in the DOLCE conceptualisation.[49]

An important methodological point must be made at this stage of the discussion. Even though the CRM has been developed by applying a precise analysis of the identity, unity and essence of classes, the methodology that explains the taxonomies is not that of OntoClean, but rather an object-oriented approach based on the analysis of properties, understood here as the expression of relations between entities. The function of the *crm:E72 Legal Object* class is thus to provide its descendant classes with the properties that associate these entities with the actors exercising a right over them (*crm:P105 right held by*) as well as with the right itself (*crm:P104 is subject to crm:E30 Right*), the latter being expressed in the form of a propositional object with no explicit connection to time. The CRM uses a multiple inheritance approach that combines within the class hierarchy both those classes that are "essential" in the OntoClean sense and those that provide additional qualifications in the form of properties, which has led to the CRM being called a "property-centric ontology".[50] Properties are understood in the context of CRM in the sense of relationships, not of essential characteristics of entities.

---

48  Cf. the *scope note* of the *crm:E27 Site* class: "In contrast to the purely geometric notion of E53 Place, this class describes constellations of matter on the surface of the Earth or other celestial body, which can be represented by photographs, paintings and maps", https://ontome.net/class/26 .

49  Cf. the "classification principle" in Stefano Borgo et al., DOLCE: A Descriptive Ontology for Linguistic and Cognitive Engineering, *Applied Ontology*, 17.1 (2022), https://doi.org/10.3233/AO-210259.

The reasons for choosing this approach, which combines two apparently incompatible methodologies, were clearly expressed by its creator, Martin Doerr, in an article entitled *The Dream of a Global Knowledge Network*, which not only presents the CRM as a "nearly generic information model", but on the basis of this approach paves the way for the realisation of the project of interoperability and scalability of information reuse that we presented in the introduction.[51] This choice has proved to be effective in terms of achieving the interoperability goals envisaged by the CRM in the field of museum data integration, but at the same time a foundational ontological analysis allows us to identify the limitations and aspects to be completed of the CRM, especially if we want to use it to conceptualise the domain of research in HSS.

Among the most relevant questions in terms of indispensable complements, let us retain that of the treatment of the properties of the objects, understood in the sense of *Quality* as defined in DOLCE. Let us note beforehand that the notion of *crm:E2 Temporal Entity* covers all the phenomena that take place in a limited period of time, with an explicit reference to the notion of Perdurant used by DOLCE. A careful analysis of this CRM class, from a property-centred perspective, shows that indeed all its properties express either a temporal relation to other phenomena —in the sense of Allen's temporal properties[52]— or a relation to a *crm:E52 Time-Span* whose function is to establish a specific position in the abstract time frame. Let us also note that, despite the identity of the name, the ontological essence of the *TemporalEntity* class of the *Time ontology* in OWL[53] is not the same as that of *crm:E2 Temporal Entity*, but much more that of *crm:E52 Time-Span*, because they actually express a temporal region in the sense of DOLCE, whereas *crm:E2 Temporal Entity* represents a phenomenon that can be observed or even photographed.

Subclasses of *crm:E2 Temporal Entity* include *crm:E4 Period*, which is the root of the conceptualisation of all physical or cultural events, and *crm:E3 Condition State*, which has been interpreted in the sense of phase, but could in fact be understood as a class equivalent to DOLCE's *Quality*, the absence of which in the CRM has been noted. The only corresponding class seems to be *crm:E16 Measurement*, which uses the *crm:E54 Dimension* class to represent a region in a quantitative abstract space defined by a unit of measurement. Note that the phenomenon captured by the class *crm:E16 Measurement* is the activity of observation, e.g. finding the length of a bridge on a given date. The choice to limit the conceptualized information to the observation activity, and to exclude the objects' intrinsic qualities from the model has the significant epistemological consequence of restricting the CRM —in this respect— to the perspective of factoïds, because we will have to inform several times in the information system the same length that this bridge was measured at different times, whereas the aggregated factual information that one would like to have for research purposes —i.e. the fact that such and such a bridge had this particular length during a given time-span, before it was transformed in such and such a year— is excluded from the CRM conceptualisation on principle.

---

50   Martin Doerr , The CIDOC Conceptual Reference Module: An Ontological Approach to Semantic Interoperability of Metadata, *AI Magazine* 24.3 (2003), 75-92, https://doi.org/10.1609/aimag.v24i3.1720.
51   Doerr Martin, Dolores Iorizzo, The dream of a global knowledge network—A new approach, *Journal on Computing and Cultural Heritage* 1.1 (2008), 1-23, https://doi.org/10.1145/1367080.1367085.
52   J. Holmen, Christian-Emil Ore, Deducing Event Chronology in a Cultural Heritage Documentation System, in *Making History Interactive. Computer Applications and Quantitative Methods in Archaeology (CAA),* Frischer, B., J. Webb Crawford and D. Koller (eds.), (Oxford: Arcaeopress, 2010), 122–29.
53   https://www.w3.org/TR/owl-time/#time:TemporalEntity.

It therefore seems advisable to add the *sdh:C1 Entity Quality* class to the SDHSS extension, which corresponds to DOLCE's notion of a *time-indexed quality* and makes it possible to add an essential component to the conceptualisation of research in HSS. Indeed, it will be possible to deal with both qualitative and quantitative qualities of objects, and their evolution over time, in a different and complementary way to the events that structure the CRM. The class *sdh:C1 Entity Quality* is defined as a subclass of *crm:E2 Temporal Entity* because it has the same substance as the latter, i.e. being an observable phenomenon limited in time, to which it adds the peculiarity of being a qualitative or quantitative property inseparable from the object it qualifies.

This not only conforms to the notion of compatibility with the CRM in a logic of specialisation of classes and inheritance of properties —the temporal ones in this case— but also takes account of the fact that, according to DOLCE's analysis, perdurants subsist by virtue of the relation of participation of endurants, which in the CRM occurs only at the level of the *crm:E5 Event* class: it is only at this level of the class hierarchy that actors with the property *crm:P11 had participant* and other objects with the property *crm:P12 occurred in the presence of* are associated. In fact, physical objects are also virtually present in the *crm:E4 Period* class, which introduces the projection into physical space with the *crm:P8 property took place on or within* a *crm:E18 Physical Thing*, but explicitely only in the *crm:E5 Event* class.

The qualities of SDHSS are thus on the same hierarchy level as the events of CRM —the perdurants of DOLCE— and, as subclasses of *crm:E2 Temporal Entity*, they are conceptualised as phenomena located in time but without direct reference to physical space. Two properties, *sdh:P8 effects* and *sdh:P9 ends*, associate events in the spatio-temporal world with the qualities they initiate or terminate. If no property directly associates the class *sdh:C1 Quality* with the objects they are related to, this is because qualities are not the same for different objects' classes, and it is therefore more appropriate, according to the object-oriented modeling methodology, to introduce subclasses of *sdh:C1 quality* in order to conceptualize specific qualities according to different types of objects. The *sdh:C1 Quality* class is indeed a powerful component of the extension because it allows to conceptualize many of the properties of objects that appear as time-bound phenomena and, as such, are inexpressible in the "event-centric" approach of CRM.

## *Intentionality and social life*

This is particularly true for a foundational approach of mental and social life, as these are at the root of most of the phenomena studied by the HSS. The CRM limits its analysis of these phenomena to what is expressed in materiality: "What goes on in our minds or is produced by our minds is also regarded as part of the material reality, as it becomes materially evident to other people at least by our utterances, behavior and products".[55] In other words, CRM accounts for social phenomena only by modelling their manifestation in 'materiality', i.e. in observable spatio-temporal events. It is in this sense that the classes *crm:E66 Formation* or *crm:E68 Dissolution*, which deal with the beginning and end of the existence of groups, or *crm:E85 Joining* and *crm:E86 Leaving*, which express the relationships of actors with groups, are to be intended. These classes are conceptualised as projections of an intentional reality into the world of spatio-temporal events because the CRM precludes itself from modelling intentional reality as such: a class that expresses a person's membership of a group during a given period is therefore, in principle, excluded from the domain of CRM discourse because it is not directly observable in materiality. How can we then deal with the political roles of people, the legal domiciles of companies, in a word: the complex properties of objects that result from social phenomena that exist only in the representations of people?

---

55  Definition of the CIDOC Conceptual Reference Model, https://cidoc-crm.org/Version/version-7.1.2.

The SDHSS extension introduces the class *sdh:C4 Intention* as a subclass of *sdh:C1 Entity Quality*, in order to integrate intentionality as envisaged by social philosophy as well as social psychology and sociology, based on the notion of mental *representations*, individual or collective.[56] This notion is conceptualised in accordance with a widespread understanding in these disciplines — formulated in a particularly precise way by the philosopher John Searle— which observes that people, individually or in groups, pay attention to objects through their own representations.[57] In the logic of the epistemological approach presented above, the conceptualisation of the class *sdh:C4 Intention* therefore does not intervene in the philosophical debate or in the scientific explanation of this phenomenon, but confines itself to constructing a concept that captures a central aspect of the foundation of the social sphere, leaving it to the various scientific disciplines to define and explain it.

Intentionality is thus conceived as a quality inherent to the mind of a person, or of several persons in a logic of collective intentionality, who mentally adhere to (shared) representations about an object. The proposed model avoids entering into the epistemological debate about the existence of a collective intentionality that goes beyond the sum of individual ones, and limits itself to modelling the existence of instances identified by the class *sdh:C9 Intentional Entity* —be they humans individually or in groups, animals or digital artefacts— that are capable of making a classification about an object at a given moment, i.e. a connotation that provides the object with a particular meaning in the context of representations expressed as instances of the *crm:E89 Propositional Object* class.[57] Intentionality is thus conceptualised as a quality of the human brain, while this human organ physically and biologically supports individual or collective representations (Fig. 4). This mental, individual and collective world underlies social life and makes it possible to account for phenomena such as the attribution of roles to persons, the ownership of objects, the membership of groups, etc., whose reality is not inherent in the objects concerned (persons or objects), but exists by virtue of a quality of the observers' brains and, of course, of the persons concerned. In this way, the ontology can deal with the fact that, in the same country and at the same time, two different groups of observers may or may not consider a particular person to have been legitimately elected president.

---

56  *E.g.: Encyclopedia of critical psychology,* Thomas T. (ed.) (New York: Springer Reference, 2014) (especially entries: Interobjectivity; Social Constructionism; Social Representations; Socialization); *The Cambridge Handbook of Social Representations,* Gordon Sammut et al., eds. (Cambridge: University Press, 2015).
57  John Searle, *Making the Social World: The Structure of Human Civilization* (Oxford: University Press, 2010), https://doi.org/10.1093/acprof:osobl/9780195396171.001.0001.
57  https://plato.stanford.edu/entries/intentionality/ ; https://plato.stanford.edu/entries/collective-intentionality/.

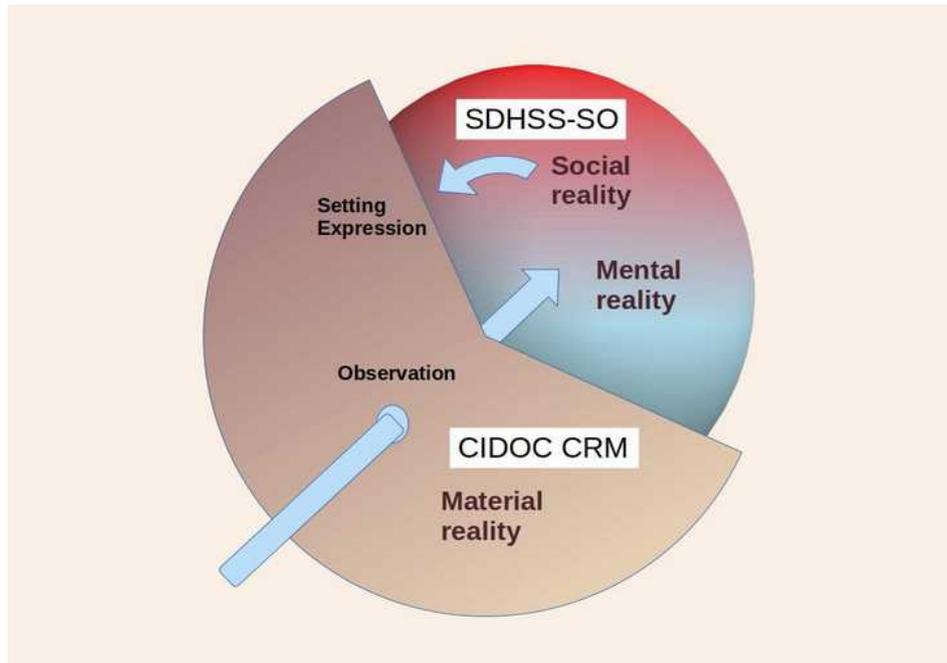

Figure 4. Material, mental and social reality

The conceptualisation adopted here is inspired by, and fits into, the ontological analysis of the *D&S* ontology around the class *Situation,* presented above, that is conceived as a specific and virtually discordant interpretation of the same world events, a conceptualisation developed in a constructivist perspective around the notion of *intentional collective.*[58] The class *sdh:C4 Intention* thus captures the information produced by the observation of social phenomena and becomes the root of a multitude of subclasses —in different extensions of the lower level of abstraction— acquiring a position equivalent to the class *crm:E5 Event* in the SDHSS class taxonomy. The coherence between the intentional level and the level of physical materiality that grounds the CRM ("material reality is regarded as whatever has substance that can be perceived with senses or instruments"[59]) is established by the property *sdh:P43 has setting*, which associates the mental phenomenon with its substratum located in the sphere of spatio-temporal phenomena (Fig. 4). For example, the intentional phenomena provoked in the mind of the reader by the reading of this text are realised by the fact that the eyes scan the signs and the neurons interpret them, whether the reader is sitting, standing, walking or all three in succession, provided that the medium on which this instance of the class *crm:E73 Information Object* is located is held in the hands and that the language in which it is written understood. These physical and intentional phenomena are complementary and inseparable, but distinct.

In the SDHSS core extension of CRM, two subclasses of *sdh:C4 Intention* express the two sides, static and dynamic, of intentional and social phenomena. On the one side, the *sdh:C7 Intentional State* class comprises opinions, beliefs, certainties, doubts, etc. of a person or collective of persons about representations concerning one or more objects. The representations, present in the mind in the form of propositional objects, are thus considered to be stable over a given period of time. This class is further specialized by the *crm:C30 Connotation* class, which expresses the time-indexed classification of entities with individually or socially defined types, and is the root of all classes expressing classifications, roles, legal connotations, etc.

---

58  Gangemi, Norms and plans (see fn. 27).
59  https://cidoc-crm.org/Version/version-7.1.2, "Modelling principles".

On the other side, the *sdh:C10 Intentional Event* class expresses the dynamic dimension of intentional phenomena, i.e. the changes of mind of persons or human collectives, or other dynamic phenomena taking place in the mind, such as reading or conversation, which are likely to bring about a change in intentional states. Intentional events, insofar as they happen in the minds of one or more persons, do not directly have a projection in geographical space. But as they are always related to sets of human actions that provide the setting for the given intentional events (*sdh:P43 is setting for*) they share with these physical events the geographical location. E.g. having a conversation with one or more persons has the correspondent physical events of sitting in a coffee house as its setting.

A relevant form of intentional events is represented by speech acts and other forms of intentional action aimed at influencing representations about things. The *sdh:C46 Intentional Expression* class includes expressions of intentionality by means of speech acts, writings, actions or activities. Representations about objects (the substance of intention) are expressed by intentional entities to other intentional entities in order to communicate their interpretations of reality, to express wishes, to give orders, to make social roles existing, etc. Raising a hand to vote involves two distinct events, both intentional and physical.[62] The aforementioned classes *crm:E65 Creation*, *crm:E66 Formation, crm:E85 Joining, crm:E86 Leaving* are all to be considered —depending on the circumstances— as subclasses of the *sdh:C10 Intentional Event* or *sdh:C46 Intentional Expression* classes, since they certainly involve a set of activities in the spatio-temporal world in order to be perceived by other humans, but fundamentally belong to the intentional mental world and therefore cause changes in representations about groups, people who belong to them, their social roles, etc. (Fig. 4).

While the aforementioned classes provide a foundational approach to social life, further classes and properties describing more specific aspects are needed, and proposed in additional extensions to the SDHSS ontology ecosystem, in line with the layered methodology adopted. At a more general level, the *Social, Legal and Economic Life* namespace (prefix: sdh-so) allows information about basic aspects of social life to be expressed. This can be illustrated in relation to some of the issues raised above. The class *sdh-so:C27 Legal Connotation* models the fact that a persistent item is perceived as having a legal quality by human groups, a society or parts of it, over a given period of time. Legal is intented here in a very general sense, in the context of any kind of more or less defined custom or law: the *sdh-so:C17 Custom or Law* class collects a set of rules that apply to a group and its members, and that define the rights and duties of the actors involved, as well as the rights and liabilities regarding specific things, such as ownership. The *sdh-so:C27 Legal Connotation* class thus expresses the same information content as the *crm:E30 Right* class, but without the need to introduce a *crm:E72 Legal Object* class into the persistent item taxonomy, just by adding a socially defined, time-indexed connotation to objects whose own substance is defined by their rigid properties, in the sens of OntoClean.

---

62  John Searle, *Making the Social World* (see fn. 56), 33-35.

In discussing the conceptualisation proposed in the ontology of the Early Modern Professorial Career Patterns project, we raised the issue of modelling temporary conditions or social roles such as student, lecturer, professor, as well as expressing membership in academies and holding tenure as distinct from actual teaching activity. The foundational approach presented in this chapter allows to conceptualise the relevant distinctions to be applied to the subclasses of the root class *pcp:StageOfLife*, separating those that model spatio-temporal events, such as birth or travel, and those that express social roles. The *sdh-so:C13 Social Role Embodiment* class, a subclass of *sdh-so:C27 Legal Connotation*, models the fact of having a more or less formally defined social role or function in a group during a given time-span, for instance being lecturer or professor at a university. This phenomenon of social connotation only exists in collective intentionality, but it provides humans with (socially) real deontic power, i.e. prerogatives and rights that are exercised in relation to other persons and groups.[63] The information about holding a chair can thus be modelled as distinct from a teaching activity, and also from the mention of it in the yearly planned course outlines, which are only the traces of the past reality that allow historians to observe and model these social intentional phenomena in their own right.

## *Conclusion*

At the end of this epistemological and semantic analysis, three elements seem relevant to keep in mind. Firstly, talking about the interoperability of HSS research data presupposes replacing data collection within the context of an analysis of knowledge production. The most relevant and useful content of research digital data in view of their reuse, according to the FAIR principles, is *information*, intended as a representation of objects, their properties and relationships. Different disciplines and research projects will be interested in different aspects of reality, in different objects, from different perspectives and with different emphasis. However, if we rigorously apply the indispensable separation between two distinct phases of research, one producing the digital data as a vehicle for the most objective information possible, and the other introducing the encodings that allow analysis, we will obtain a rich universe of reusable information that will allow to represent multiple facets of reality in a cumulative graph of increasing volume and quality.

Secondly, this project can only be realised by applying established methods of ontological analysis, in particular by using foundational ontologies and by distinguishing different levels of abstraction in order to jointly develop an ecosystem of shared and reusable ontologies. The online application *ontome.net* has been designed as a support to facilitate the implementation of this vision, allowing different projects to adopt data models specific to their research, while reusing existing ones as much as possible and inscribing them in an ontology articulated in different namespaces extending the core level to more specific research subdomains.

Thirdly, it seems sensible to adopt the CIDOC CRM as a core ontology providing the high-level classes needed to describe factual, spatio-temporal information in the HSS domain. It is essential, however, to increment it with a high-level extension, Semantic Data for Humanities and Social

---
63   John Searle, *Making the Social World* (see fn. 56), 145ff.

Sciences (SDHSS), in order to cover the entire domain of research of HSS and furthermore add subdomain extensions, at different levels of abstractions, that are needed for more specific research agendas. This coherent ecosystem of ontologies will provide HSS with a range of reusable conceptualisations to ensure interoperability that is much richer semantically than the simple 'technical' alignment of ontologies and much less costly in terms of time and resources than having to reinvent a conceptualisation for each project.

This vision and methodological approach is intended to promote the application of FAIR principles to HSS research and to enable the creation of a giant information graph for these disciplines. It remains to be seen whether the research communities will be able to open up to this transition, which is both epistemological and practical. To succeed, it will require a new form of collective commitment that transcends disciplinary boundaries and rejects individualist logics that are impervious to the vision of the FAIR principles. It also represents a civic commitment of the HSS to take a stand against the economic and symbolic power of the web giants, based in particular on inaccessible knowledge graphs oriented towards financial profitability. A giant, distributed information graph, collaboratively maintained by HSS research, would make it possible to defend a critical and humanist analysis of the realities of the world.